\documentclass[fleqn]{rendiconti}

\usepackage{amsmath}
\usepackage{xspace}
\usepackage{url}
\usepackage{xcolor}
\usepackage{soul}
\usepackage{amsfonts}
\usepackage{amsmath}
\usepackage{amssymb}
\usepackage{z-eves}
\usepackage[latin1]{inputenc}
\usepackage{tikz}
\usepackage{framed}
\usepackage{comment}
\usetikzlibrary{positioning,patterns,shapes,arrows,shadows,backgrounds,fit}

\newcommand{\setlog}{$\{log\}$\xspace}
\newcommand{\INT}{\textsf{INTERVAL}\xspace}
\newcommand{\CLPSET}{CLP($\mathcal{SET}$)\xspace}

\newcommand{\SET}{\mathcal{SET}}
\newcommand{\BR}{\mathcal{BR}}
\newcommand{\RIS}{\mathcal{RIS}}
\newcommand{\TX}{\mathcal{X}}
\newcommand{\CARD}{\card{\cdot}}

\newcommand{\LSET}{\mathcal{L}_{\SET}}

\newcommand{\LBR}{\mathcal{L}_{\BR}}
\newcommand{\LRIS}{\mathcal{L}_\mathcal{RIS}}
\newcommand{\LCARD}{\mathcal{L}_{\CARD}}

\newcommand{\SSet}{\mathsf{S}}
\newcommand{\SInt}{\mathsf{Z}}
\newcommand{\Ur}{\mathsf{U}}

\newcommand{\SVar}{\mathcal{V}}
\newcommand{\FSet}{\mathcal{F}_\SSet}
\newcommand{\FUr}{\mathcal{F}_\Ur}
\newcommand{\FInt}{\mathcal{F}_\SInt}
\newcommand{\PiInt}{\Pi_\SInt}
\newcommand{\PiSet}{\Pi_\mathsf{S}}
\newcommand{\PiRB}{\Pi_\mathit{R}} %{\Pi_\mathit{RB}}

\newcommand{\disj}{\parallel}

\newcommand{\p}{(\cdot,\cdot)}   %pair
\newcommand{\card}[1]{\lvert #1 \rvert}

\newcommand{\true}{\mathit{true}}
\newcommand{\false}{\mathit{false}}

\renewcommand{\plus}{\mathbin{\scriptstyle\sqcup}}
\newcommand{\Cp}{\mathsf{cp}}
\newcommand{\es}{\{\}}
\newcommand{\set}[2]{\{#1 \plus #2\}}

\newcommand{\flt}{\phi}   % Filter
\newcommand{\ptt}{u}      % Pattern
\renewcommand{\Cup}{\mathsf{un}}
\newcommand{\Ncup}{\mathsf{nun}}
\renewcommand{\Cap}{\mathsf{inters}}
\newcommand{\In}{\mathbin{\mathsf{in}}}
\newcommand{\Nin}{\mathbin{\mathsf{nin}}}
\newcommand{\Neq}{\mathbin{\mathsf{neq}}}
\newcommand{\Disj}{\mathsf{disj}}
\newcommand{\Dom}{\mathsf{dom}}
\newcommand{\Ran}{\mathsf{ran}}
\newcommand{\Subseteq}{\mathsf{subset}}
\newcommand{\Comp}{\mathsf{comp}}
\newcommand{\Inv}{\mathsf{inv}}
\newcommand{\Oplus}{\mathsf{oplus}}
\newcommand{\Apply}{\mathsf{apply}}
\newcommand{\ApplyTo}{\mathsf{applyTo}}
\newcommand{\Pfun}{\mathsf{pfun}}
\newcommand{\Npfun}{\mathsf{npfun}}
\newcommand{\Or}{\mathbin{\mathsf{or}}}
\newcommand{\Ris}{\mathsf{ris}}
\newcommand{\Ncomp}{\mathsf{ncomp}}
\newcommand{\Is}{\mathbin{\mathsf{is}}}
\newcommand{\Foreach}{\mathsf{foreach}}
\newcommand{\Size}{\mathsf{size}}

\newcommand{\Id}{\mathsf{id}}
\newcommand{\Set}{\mathsf{set}}
\newcommand{\Nset}{\mathsf{nset}}
\newcommand{\Rel}{\mathsf{rel}}
\newcommand{\Nrel}{\mathsf{nrel}}
\newcommand{\Pair}{\mathsf{pair}}
\newcommand{\Npair}{\mathsf{npair}}
\newcommand{\Int}{\mathsf{int}}
\newcommand{\Nint}{\mathsf{nint}}
\newcommand{\Integer}{\mathsf{integer}}

\newcommand{\why}[1]{\tag*{{\footnotesize [by #1]}}}
\newcommand{\ie}[1]{\tag*{{\footnotesize [#1]}}}

\newcommand{\hlg}[1]{\sethlcolor{green}\hl{#1}}

\RendicontiPagina{1}{M. Cristi\'a and G. Rossi}{\setlog: Set Formulas as Programs}
\def\DOI{xxxxx} % DO NOT MODIFY THIS FIELD

\title{\setlog: Set Formulas as Programs}

\author{Maximiliano Cristi\'a and Gianfranco Rossi}
\summary{%
\setlog is a programming language at the intersection of Constraint Logic Programming, set programming and declarative programming. But \setlog is also a satisfiability solver for a theory of finite sets and finite binary relations. With \setlog programmers can write abstract programs using all the power of set theory and binary relations. These programs are not very efficient but they are very close to specifications. Then, their correctness is more evident. Furthermore, \setlog programs are also set formulas. Hence, programmers can use \setlog again to automatically prove their programs verify non trivial properties. In this paper we show this development methodology by means of several examples.
}
\date{Month dd, yyyy\\ Revised Month dd, yyyy\\Accepted Month dd, yyyy}
\intesta

\begin{document}
\maketitle

{MSClassification, separated, by, commas}

\section{Introduction}
%[da Apt] Formulas as programs is argued to yield a realistic approach to
%programming that has been realized in the implemented programming language Alma
%that combines the advantages of imperative and logic programming.}

In 1999 Apt and Bezem \cite{Apt1999} proposed a programming paradigm based on
the concept of \emph{formulas as programs} as an alternative approach to
\emph{formulas as types} or \emph{proofs as programs}
\cite{Bates:1985:PP:2363.2528, DBLP:journals/iandc/CoquandH88,
DBLP:books/daglib/0070531}. In the latter approach, formal proofs of properties
about the correct behavior of a program contain a Lambda calculus term (i.e., a
program) which is a correct implementation of that behavior. Hence, by proving
properties concerning the behavior of a program one gets in addition
correct programs for free. In the 'formulas as programs' approach, the formula
(specification) is itself a program. No formal proof is needed to get a
program, but the specification might not verify some desired properties making
the program faulty. Clearly, both approaches have their advantages and
disadvantages. For example, in the 'proofs as programs' approach one have the
first version of the program after performing a formal proof of some property
and extracting the Lambda term (which is not always easy), but this first
version is correct by construction. On the other hand, in the 'formulas as
programs' approach one quickly have a first version of the program but it can
be wrong, although it can be improved by experimenting with it.

Set theory is deemed as a good vehicle to concisely and accurately describe
algorithms and software systems. Formal specification languages such as Z
\cite{Spivey00}, B \cite{schneider2001b}, TLA+ \cite{DBLP:conf/asm/2018} and
VDM \cite{DBLP:books/daglib/0067105} support this claim. In this paper we show
how set-based specifications can be made to fit in the `formulas as programs'
paradigm.

\setlog is a constraint logic programming (CLP) language which provides the
fundamental forms of set designation, along with a number of basic operations
for manipulating them, as first-class entities. Various new features have been
added to the core part of the language since the initial development of \setlog
\cite{DBLP:journals/jlp/DovierOPR96}. Among them, basic facilities for
representing and manipulating integer expressions (integrating the CLP(FD) and
CLP(Q) solvers), binary relations, partial functions, Cartesian products and
restricted intensional sets. But \setlog is also a \emph{satisfiability
solver}. The CLP language and the satisfiability solver are two sides of the
same and only system. That is, \setlog is not the integration of a CLP
interpreter with a satisfiability solver; instead, it is based on mathematical
and computational models that produce such a tool.

This means that a piece of \setlog code is both \emph{a program and a formula}.
We call this the \emph{program-formula duality}. Therefore, when \setlog
programmers write code they are writing both a program and a formula. In other
words, they are writing a program \emph{as a} formula. When seen as a program,
programmers can execute it; when seen as a formula, they can
\emph{automatically} prove properties true of it. Hence, a \setlog programmer
writes some code and execute it to see how it works. If everything goes right,
(s)he can use \setlog again to automatically prove properties of that program.
All with the same and only formal text and with the same and only tool. Once a
\setlog program is shown to verify some property, we can be sure that all of
its executions are correct with respect to that property.

However, \setlog has some limitations. \setlog programs perform poorly compared
with logic, functional or imperative programs. We see \setlog programs as
\emph{functional prototypes} or \emph{executable specifications}. Not every
property true of a \setlog program can be automatically proved with \setlog.
Further, proving some properties may take too much computing time making the
process unpractical. The capacity of \setlog in automatically proving
properties depends on the program-formula fitting inside of the decision
procedures implemented by the tool.

In this paper we will show this program-formula duality through some
revealing examples.

%Instead of using {log}'s concrete syntax we will use a more compact and
%Latex--oriented form which nonetheless can be mechanically translated into
%{log}--the concrete programs can be found in the supplemental material.

%%%%%%%%%%%%%%%%%%%%%%%%%%%%%%%%%%%%%%%%%%%%%%%%%%%%%%%%%%%%%%%%%%%%%
\section{\label{setlog}\setlog}
%%%%%%%%%%%%%%%%%%%%%%%%%%%%%%%%%%%%%%%%%%%%%%%%%%%%%%%%%%%%%%%%%%%%%

\setlog is a publicly available satisfiability solver and a set-based,
constraint-based programming language implemented in Prolog \cite{setlog}.

\setlog implements a decision procedure for the theory of \emph{hereditarily
finite sets} ($\SET$), i.e., finitely nested sets that are finite at each level
of nesting \cite{Dovier00}; a decision procedure for a very expressive fragment
of the theory of finite set relation algebras ($\BR$)
\cite{DBLP:journals/jar/CristiaR20,DBLP:conf/RelMiCS/CristiaR18}; a decision
procedure for the theory of finite sets with restricted intensional sets
($\RIS$) \cite{DBLP:conf/cade/CristiaR17,Cristia2021}; a
decision procedure for the theory of hereditarily finite sets extended with
cardinality constraints ($\CARD$); a decision procedure for the latter extended
with integer intervals ($\CARD$); and uses Prolog's CLP(Q) to provide a
decision procedure for the theory of integer linear arithmetic
\cite{DBLP:conf/cp/HolzbaurMB96}. All these procedures are integrated into a
single solver, implemented in Prolog, which constitutes the core part of the
\setlog tool. Several in-depth empirical evaluations provide evidence that
\setlog is able to solve non-trivial problems
\cite{DBLP:journals/jar/CristiaR20,DBLP:conf/RelMiCS/CristiaR18,DBLP:conf/cade/CristiaR17,CristiaRossiSEFM13};
in particular as an automated verifier of security properties
\cite{DBLP:journals/jar/CristiaR21,DBLP:journals/corr/abs-2009-00999}.

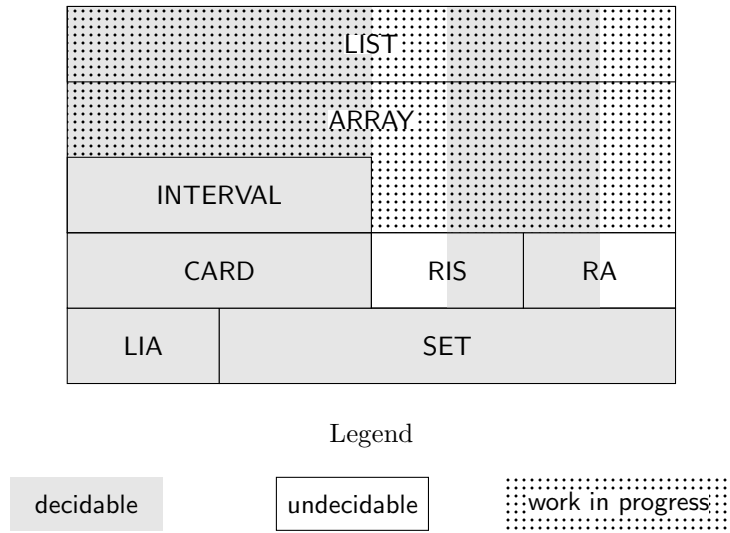
\begin{figure}
\begin{center}
\begin{tikzpicture}
  [every node/.style={transform shape},font=\sffamily,
   rectangle,text centered,minimum width=2cm,minimum height=1cm]
    %%%%%%%%%%%%%%%%%%%%%%%%%%%%%%%%%%%%%%%%%%%%%%%%%%%%%%%%%%%%%%%%%%
    %%%%%%%%%%%%%%%%%%%%%%%%%% NODES %%%%%%%%%%%%%%%%%%%%%%%%%%%%%%%%%
    %%%%%%%%%%%%%%%%%%%%%%%%%%%%%%%%%%%%%%%%%%%%%%%%%%%%%%%%%%%%%%%%%%
\draw
 (0,0) node[draw,fill=gray!20] (ilp) {LIA}
 (4,0) node [draw,minimum width=6cm,fill=gray!20] (set) {SET}
 (1,1) node [draw,minimum width=4cm,fill=gray!20] (card) {CARD}
 (4.5,1) node [minimum width=1cm,fill=gray!20] {}
 (4,1) node [draw] (ris) {RIS}
% (5,1) node [fill=white,minimum width=0.5cm] {}
 (5.5,1) node [minimum width=1cm,fill=gray!20] {}
 (6,1) node [draw] (ra) {RA}
 (1,3) node [minimum width=4cm,fill=gray!20] {}
% (1,3) node [draw,minimum width=4cm] {}
 (1,3) node [minimum width=4cm,pattern=dots] {}
 (5,3) node [minimum width=2cm,minimum height=3cm,fill=gray!20] {}
 (1,4) node [minimum width=4cm,fill=gray!20] {}
 (1,2) node [draw,minimum width=4cm,fill=gray!20] (int) {INTERVAL}
 (5,2.5) node
   [minimum width=4cm, minimum height=2cm, align=right, pattern=dots] {}
 (3,3) node [draw,minimum width=8cm,minimum height=3cm] {}
 (3,3) node [fill=white,minimum width=0cm,minimum height=0cm, inner sep=0pt] (arr) {ARRAY}
 (3,4) node [draw,minimum width=8cm,pattern=dots] {}
 (3,4) node [fill=white,minimum width=0cm,minimum height=0cm, inner sep=0pt] (list) {LIST};
\end{tikzpicture}
\end{center}

\centerline{Legend}
\begin{center}
\begin{tikzpicture}
  [every node/.style={transform shape},font=\sffamily,
   rectangle,text centered,minimum width=2cm]
\draw
 (0,0) node[fill=gray!20,minimum height=20pt] (dec) {decidable}
 (3.5,0) node[minimum height=20pt,draw] (undec) {undecidable}
% (3.5,-0.65) node {\footnotesize{$R \subseteq R \comp S$ or $R \subseteq S \comp R$}}
 (7,0) node[minimum height=20pt,pattern=dots,minimum width=3cm] {}
 (7,0) node[fill=white, inner sep=0pt] {work in progress};
\end{tikzpicture}
\end{center}
\caption{\label{fig:stack}The stack of theories dealt with by \setlog}
\end{figure}

Figure \ref{fig:stack} schematically describes the stack of the first-order
theories supported by \setlog. The fact that a theory $T$ is over a theory $S$
means that $T$ extends $S$. E.g., \textsf{CARD} extends both \textsf{LIA} and
\textsf{SET}. Figure \ref{fig:theories} shortly describes the considered
theories, showing for each of them the main constant, function and predicate
symbols. The precise definition of the first-order logic languages on which the
theories are based on are given in Appendix A.

\begin{figure}
\begin{framed}
\begin{itemize}
  \item
  \textsf{LIA}: Linear Integer Arithmetic (i.e., the theory that allows
  inequalities over sums of constant multiples of variables).\\
  Symbols: $\langle\mathsf{Z},+,*,-,=,\leq \rangle$, where $\mathsf{Z}$
  is the set of integer constants.
  \item
  \textsf{SET}: Hereditarily finite hybrid untyped extensional sets.\\
  Symbols: $\langle\Ur,\{\},\{\cdot \plus \cdot\},=,\neq,\in,\notin,\cup,\parallel
  \rangle$, where $\Ur$ is the set of ur-elements, i.e., non-set objects
  %objects which contain no elements but are distinct from the empty set,
  %(also known as atoms or individuals)
  that are used as set elements, and $\{\cdot \plus \cdot\}$ is a binary function
  symbol which serves as the extensional set constructor.
  %$\{x \plus A\}$ intuitively represents the set $\{x\} \cup A$.
  %
  \item
  \textsf{CARD}: Hereditarily finite hybrid untyped extensional sets
  with cardinality.\\
  Symbols: $\langle\mathsf{Z},+,*,-,=,\leq \rangle$,
  $\langle\Ur,\{\},\{\cdot \plus
  \cdot\},=,\neq,\in,\notin,\cup,\parallel,|\cdot|\rangle$.
  %\num
  %
  \item
  \textsf{RIS}: Hereditarily finite hybrid untyped extensional and intensional sets.\\
  Symbols: $\langle\Ur,\{\},\{\cdot \plus
  \cdot\},\{\cdot|\cdot@\cdot\},=,\neq,\in,\notin,\cup,\parallel\rangle$, where
  $\{\cdot|\cdot@\cdot\}$ is a ternary function symbol which serves as
  the intensional set constructor.
  %$\{x:D|\flt@\ptt\}$ intuitively represents the set of terms $\ptt$ such that
  %$x$ belongs to $D$ and the formula $\flt$ holds for $x$.
  %
  \item
  \textsf{RA}: Finite set relation algebras over discrete universe.\\
  Symbols: $\langle\Ur,\{\},\{\cdot \plus \cdot\},
  (\cdot,\cdot),=,\neq,\in,\notin,\cup,\parallel,\id,\comp,{}^\smile \rangle$
  \item
  \INT: Hereditarily finite hybrid untyped extensional sets and integer
  intervals with cardinality.\\
  Symbols: $\langle\mathsf{Z},+,*,-,=,\leq \rangle$,
  $\langle\Ur,\{\},\{\cdot \plus \cdot\},[\cdot,\cdot],=,\neq,\in,\notin,\cup,\parallel,|\cdot| \rangle$
  \item
  \textsf{ARRAY}: Arrays encoded as binary relations.\\
  Symbols: $\langle\mathsf{Z},+,*,-,=,\leq \rangle$, $\langle\Ur,\{\},\{\cdot \plus
\cdot\},\{\cdot|\cdot@\cdot\},[\cdot,\cdot],(\cdot,\cdot),=,\neq,\in,\notin,\cup,\parallel,|\cdot|,\id,\comp,{}^\smile
\rangle$
  \item
  \textsf{LIST}: Lists encoded as binary relations.\\
  Symbols: $\langle\mathsf{Z},+,*,-,=,\leq \rangle$, $\langle\Ur,\{\},\{\cdot \plus
\cdot\},\{\cdot|\cdot@\cdot\},[\cdot,\cdot],(\cdot,\cdot),=,\neq,\in,\notin,\cup,\parallel,|\cdot|,\id,\comp,{}^\smile
\rangle$
\end{itemize}
\end{framed} \caption{\label{fig:theories}The theories dealt with by \setlog}
\end{figure}

The integrated constraint language offered by \setlog is a quantifier-free
first-order predicate language with terms of two sorts: terms designating sets
and terms designating ur-elements. Terms of either sort are allowed to enter in
the formation of \emph{set terms} (in this sense, the designated sets are
hybrid), no nesting restrictions being enforced (in particular, membership
chains of any finite length can be modeled).

Set terms in \setlog can be of the following forms:
\begin{itemize}
\item A variable is a set term; variable names start with an uppercase letter.
\item $\{\}$ is the term interpreted as the empty set.
\item $\{x / A\}$ is called \emph{extensional set} and is interpreted as
$\{x\} \cup A$; $A$ must be a set term, $x$ can be any term accepted by \setlog
(basically, any Prolog uninterpreted term, integers, ordered pairs, other set
terms, etc.).\footnote{Note that $\{\_/\_\}$ is the concrete syntax for the
(abstract) set term $\{\_\plus\_\}$ of Figure \ref{fig:theories}.}

As a notational convention, set terms of the form $\{t_1 / \{t_2 \,/\, \cdots
\{ t_n / t\}\cdots\}\}$ are abbreviated as $\{t_1,t_2,\dots,t_n / t\}$, while
$\{t_1 / \{t_2 \,/\, \cdots \{ t_n / \es\}\cdots\}\}$ is abbreviated as
$\{t_1,t_2,\dots,t_n\}$.
\item $\Ris(X \In A, \phi)$ is called \emph{restricted intensional set} (RIS)
and is interpreted as $\{x : x \in A \land \phi\}$ where $\phi$ is any \setlog
formula, $A$ must be a set term, and $X$ is a bound variable local to the RIS.
Actually, RIS have a more complex and expressive structure
\cite{DBLP:conf/cade/CristiaR17,Cristia2021}.   %
\item $\Cp(A,B)$ is interpreted as $A \times B$, i.e., the Cartesian
product between $A$ and $B$.
\item $\Int(A,B)$ is interpreted as $\{x \in \num | A \leq x \leq B\}$.
\end{itemize}

Set terms can be combined in several ways: binary relations are hereditarily
finite sets whose elements are ordered pairs and so set operators can take
binary relations as arguments; RIS and integer intervals can be passed as
arguments to set operators and freely combined with extensional sets. \setlog
is an untyped formalism; variables are not declared; typing information can be
encoded by means of constraints.\footnote{Recently, a type system and a type
checker have been added to the base language for those users who feel more
comfortable with typed formalisms.}

Set operators are encoded as atomic predicates, and are dealt with as
constraints. For example: $\Cup(A,B,C)$ is a constraint interpreted as $C = A
\cup B$. \setlog implements a wide range of set and relational operators
covering most of those used in Z. For instance, $\In$ is a constraint
interpreted as set membership (i.e., $\in$); $=$ is set equality; $\Dom(F,D)$
corresponds to $\dom F = D$; $\Subseteq(A,B)$ corresponds to $A \subseteq B$;
$\Comp(R,S,T)$ is interpreted as $T = R \comp S$ (i.e., relational
composition); and $\Apply(F,X,Y)$ is equivalent to $\Pfun(F) \And [X,Y] \In F$,
where $\Pfun(F)$ constrains $F$ to be a (partial) function.

A number of other set, relational and integer operators (in the form of
predicates) are defined as \setlog formulas, thus making it simpler for the
user to write complex formulas. Dovier et al. \cite{Dovier00} proved that the
collection of predicate symbols $\{=,\neq,\in,\notin,\cup,\parallel\}$ is
sufficient to define constraints implementing the set operators $\cap$,
$\subseteq$ and $\setminus$.
%For example, $A \subseteq B$ can be defined by the $\LCARD$ formula
%$\Cup(A,B,B)$.
This result has been extended to binary relations
\cite{DBLP:journals/jar/CristiaR20} by showing that adding to the previous
collection the predicate symbols $\{\id,\comp,{}^\smile\}$ is sufficient to
define constraints for most of the classical relational operators, such as
$\dom$, $\ran$, $\dres$, $\rres$, etc.. Similarly, $\{=,\neq,\leq\}$ is
sufficient to define $<$, $>$ and $\geq$. We call predicates defined in this
way, \emph{derived constraints}.

\begin{remark}
Establishing which predicates can be expressed as derived constraints and
which, on the contrary, cannot is a critical issue. Primitive constraints are
processed by possibly recursive \emph{ad hoc} rewriting procedures, that allow
one to implement a form of universal quantification which is not provided by
the language. Conversely, derived constraints are processed by simply replacing
them by quantifier-free first order formulas.

Choices about primitive vs. derived constraints can be different. For example,
in \cite{DBLP:conf/cav/CristiaR16} we use $\Dom$ and $\Ran$ in place of $\Inv$
and $\Id$. However, since the $\Inv$ predicate for binary relations appears not
to be definable in terms of the other primitive predicates, $\Inv$ has been
included as a primitive constraint in later work
\cite{DBLP:journals/jar/CristiaR20}, to enlarge the expressiveness of the
constraint language. At the same time, $\Dom$ and $\Ran$ are moved out of the
primitive constraints, since they turn out to be definable in terms of $\Id$,
$\Comp$ and $\Inv$, thus reducing the number of primitive constraints.

Proving that the selected collection of primitive constraints is the minimal
one, as well as comparing one choice to another in terms of, e.g., expressive
power, completeness, effectiveness, and efficiency, is a challenging issue for
future work.\qed
\end{remark}

Negation in \setlog is introduced by means of so-called \emph{negated
constraints}. For example $\Ncup(A,B,C)$ is interpreted as $C \neq A \cup B$
and $\Nin$ corresponds to $\notin$---in general, a constraint beginning with
`$\mathsf{n}$' identifies a negated constraint. Most of these constraints are
defined as derived constraints in terms of the existing primitive constraints;
thus their introduction does not really require extending the constraint
language. For formulas to fit inside the decision procedures implemented in
\setlog, users must only use this form of negation.
%Thanks to the availability of negative constraints, (general) logical negation
%is not strictly necessary in $\LCARD$.

Formulas in \setlog are built in the usual way by using conjunctions ($\&$)
and disjunctions ($\Or$) of atomic constraints.
%; they must finish with a dot (as a Prolog query).

\begin{example}\label{ex:setlogformulas}
The following are two simple formulas accepted by \setlog:
\begin{verbatim}
   a in A & a nin B & un(A,B,C) & C = {X / D}.

   un(A,B,C) & N + K > 5 & size(C,N) & B neq {}.
\end{verbatim}
\qed
\end{example}

%In \setlog sets and binary relations are first-class entities of the language.

As concerns constraint solving, the \setlog solver repeatedly applies specialized
rewriting procedures to its input formula $\Phi$ and returns either $\false$ or
a formula in a simplified form which is guaranteed to be satisfiable with
respect to the intended interpretation. Each rewriting procedure applies a few
non-deterministic rewrite rules which reduce the syntactic complexity of
primitive constraints of one kind. At the core of these procedures is set
unification \cite{Dovier2006}. The execution of the solver is iterated until a
fixpoint is reached, i.e., the formula is irreducible.

The disjunction of formulas returned by the solver represent all the concrete
(or ground) \emph{solutions} of the input formula. Any returned formula is
divided into two parts: the first part is a (possibly empty) list of equalities
of the form $X = t$, where $X$ is a variable occurring in the input formula and
$t$ is a term; and the second part is a (possibly empty) list of primitive
constraints.

%%%%%%%%%%%%%%%%%%%%%%%%%%%%%%%%%%%%%%%%%%%%%%%%%%%%%%%%%%%%%%%%%%%%%
\section{Uses of \setlog}
%%%%%%%%%%%%%%%%%%%%%%%%%%%%%%%%%%%%%%%%%%%%%%%%%%%%%%%%%%%%%%%%%%%%%

In this section we show examples on how \setlog can be used as a programming
language (\ref{programming}) and as an automated theorem prover (\ref{prover}).

\subsection{\label{programming}\setlog as a programming language}

\setlog is primarily a programming language, at the intersection of declarative
programming,  set programming \cite{DBLP:books/daglib/0067831} and constraint
programming. Specifically, \setlog is an instance of the general CLP scheme. As
such, \setlog programs are structured as a finite collection of \emph{clauses},
whose bodies can contain both atomic constraints and user-defined predicates.
%A clause can be seen as a subroutine or procedure. Clauses can receive zero or
%more arguments. The only way a clause can return a value is by means of one or
%more of its arguments. Under certain conditions clauses behave as formulas.
%That is a \setlog clause can be seen as both a program and a formula.
The following examples show the \emph{formula-program duality} of \setlog code
along with the notion of clause.
%this is the first time we mention the \emph{formula-program duality}

\begin{example}\label{ex:update}
If we want a program that updates function $F$ in $X$ with value $Y$ provided
$X$ belongs to the domain of $F$ and get an error otherwise, the \setlog code
can be the following:
\begin{gather*}
\mathsf{update}(F,X,Y,F\_,Error) \text{ :-} \\
\t1 F = \{[X,V]/F1\} \And [X,V] \Nin F1 \And
    F\_ = \{[X,Y]/F1\} \And \\
\t1 Error = ok \\
\t1 \Or \\
\t1 \Dom(F,D) \And X \Nin D \And \\
\t1 Error = err.
\end{gather*}
That is, $\mathsf{update}$
%is a clause that receives $F$, $X$ and $Y$ and
returns the modified $F$ in $F\_$ and the error code in $Error$---think of
$F\_$ as the value of $F$ in the next state. As $\And$ and $\Or$ are logical
connectives and $=$ is logical equality, the order of the `instructions' is
irrelevant w.r.t. the functional result---although it can have an impact on the
performance. Variable $F1$ is an existentially quantified variable representing
the `rest' of $F$. If the ordered
pair $[X,V]$ does not belong to $F$ then the unification between $F$ and
$\{[X,V]/F1\}$ will fail thus making $\mathsf{update}$ to execute the other
branch.

Now we can call $\mathsf{update}$ by providing inputs and waiting for outputs:
\[
\mathsf{update}(\{[setlog,5],[hello,earth],[tokeneer,model]\},
hello,world,G,E).
\]
returns:
\begin{gather*}
G = \{[hello,world],[setlog,5],[tokeneer,model]\}, E = ok \tag*{\qed}
\end{gather*}
\end{example}

As a programming language, \setlog can be used to implement set-based
specifications (e.g., Z specifications). As a matter of fact, many of such
specifications can be easily translated into \setlog (see \cite{cristia2021log}). This
means that \setlog can serve as a programming language in which a
\emph{prototype} of a set-based specification can be easily implemented. In a
sense, the \setlog implementation of a set-based specification can be seen as
an \emph{executable specification}.

%Since we see \setlog programs as prototypes we talk about \emph{simulations} or
%\emph{animations} rather than \emph{executions} when speaking about running
%them. However, technically, what we do is no more than running a program. The
%word \emph{simulation} is usually used in the context of \emph{models} (e.g.
%modeling and simulation). In a sense, our \setlog programs are \emph{executable
%models} of the user requirements. On the other hand, the word \emph{animation}
%is usually used in the context of formal specifications.
%
%In fact, as we will see, \setlog programs have features and properties usually
%enjoyed by specifications and models, which are rare or nonexistent in programs
%written in imperative (and even functional) programming languages.

\begin{remark}
A \setlog implementation of a set-based specification is easy to get but
usually it will not meet the typical performance requirements demanded by
users. Hence, we see a \setlog implementation of a set-based  specification
more as a \emph{prototype} than as a final program. On the other hand, given
the similarities between a specification and the corresponding \setlog program,
it's reasonable to think that the prototype is a \emph{correct} implementation
of the specification\footnote{In fact, the translation process can be automated
in many cases.}. \qed
\end{remark}

Then, we can use these prototypes to make an early validation of the
requirements.
%the basic idea is to provide inputs to the program, model or specification and
%observe the produced outputs or effects.
Validating user requirements by means of prototypes entails executing the
prototypes together with the users so they can agree or disagree with the
behavior of the prototypes. This early validation will detect many errors,
ambiguities and incompleteness present in the requirements and possible
misunderstandings or misinterpretations generated by the software engineers.
Without this validation many of these issues would be detected in later stages
of the project thus increasing the project costs. Think that if one of these
issues is detected once the product has been delivered it means to correct the
requirements document, the specification, the design, the implementation, the
user documentation, etc.

\subsection{\label{prover}\setlog as an automated theorem prover}

\setlog is also a \emph{satisfiability solver}.  This means that \setlog is a
program that can decide if formulas of some theory are \emph{satisfiable} or
not. In this case the theory is the theory of finite sets and binary relations,
combined with linear integer arithmetic.

Being a satisfiability solver, \setlog can be used as an automated theorem
prover. To prove that formula $\phi$ is a theorem, \setlog has to be called to
prove that $\lnot\phi$ is unsatisfiable.

\begin{example}
We can prove that set union is commutative by asking \setlog to prove the
following is unsatisfiable:
\begin{gather*}
\Cup(A,B,C) \And \Cup(B,A,D) \And C \Neq D.
\end{gather*}
As there are no sets satisfying this formula \setlog answers $\mathsf{no}$.
Note that the formula can also be written with the $\Ncup$ constraint:
$\Cup(A,B,C) \And \Ncup(B,A,C)$. \qed
\end{example}

Evaluating properties with \setlog helps to run correct simulations by checking
that the starting state is correctly defined. It also helps to \emph{test}
whether or not certain properties are true of the specification or not.
However, by exploiting the ability to use \setlog as a theorem prover, we can
\emph{prove} that these properties are true of the specification.

For instance, since $\mathsf{update}$ in  Example \ref{ex:update} is also a
formula we can \emph{automatically} prove properties true of it.

\begin{example}
If $Error$ is equal to $err$ then $X$ does not belong to the domain of $F$. In
order to prove this property we need to call \setlog on its negation:
\[
\mathsf{update}(F,X,Y,F\_,err) \And \Dom(F,D) \And X \In D.
\]
Then, \setlog answers $\mathsf{no}$ because the formula is unsatisfiable.
Further, we can prove that $\Dom(F,D) \And X \Nin D$ is equivalent to
$\Comp(\{[X,X]\},F,\{\})$, which allows us to refine $\mathsf{update}$ into a
version not computing the domain of $F$. In fact, $\Comp(\{[X,X]\},F,\{\})$ is
just a linear iteration over all the elements of $F$. Then, we need to
discharge the following proof obligation:
\[
\Dom(F,D) \And X \Nin D \iff \Comp(\{[X,X]\},F,\{\})
\]
by proving that its negation is unsatisfiable:
\begin{gather*}
\Dom(F,D) \And X \Nin D \And \Ncomp(\{[X,X]\},F,\{\})\tag{$\implies$}\\
\Comp(\{[X,X]\},F,\{\}) \And \Dom(F,D) \And X \In D \tag{$\Leftarrow$}
\end{gather*}
Hence, now we can write $\mathsf{update}$ as follows:
\begin{gather*}
\mathsf{update}(F,X,Y,F\_,Error) \text{ :-} \\
\t1 F = \{[X,V]/F1\} \And [X,V] \Nin F1 \And
    F\_ = \{[X,Y]/F1\} \And \\
\t1 Error = ok \\
\t1 \Or \\
\t1 \Comp(\{[X,X]\},F,\{\}) \And \\
\t1 Error = err. \tag*{\qed}
\end{gather*}
\end{example}

Furthermore, \setlog can be used to automatically discharge verification
conditions in the form of invariants. Precisely, in order to prove that an
operation $T$ preserves the state invariant $I$ we have to discharge the
following proof obligation:
\begin{equation}\label{e:inv}
I \land T \implies I'
\end{equation}
If we want to use \setlog to discharge \eqref{e:inv} we have to ask \setlog to
check if the negation of \eqref{e:inv} is \emph{unsatisfiable}. In fact, we
need to execute the following \setlog \emph{program}:
\begin{equation}\label{e:neginv}
I \land T \land \lnot I'
\end{equation}
because $\lnot(I \land T \implies I') \equiv \lnot(\lnot(I \land T) \lor I')
\equiv I \land T \land \lnot I'$.

\begin{example}
An invariant property of $\mathsf{update}$ is that $F$ is a function.
Formally, we can prove the following:
\[
\Pfun(F) \And \mathsf{update}(F,X,Y,F\_,E) \implies \Pfun(F\_)
\]
as always by proving that its negation is unsatisfiable:
\begin{equation}
\Pfun(F) \And \mathsf{update}(F,X,Y,F\_,E) \And \Npfun(F\_) \tag*{\qed}
\end{equation}
\end{example}

As these examples show, \setlog is a programming and proof platform  exploiting
the program-formula duality within the theory of finite sets and binary
relations.

In particular, many Z specifications can be easily translated into \setlog (see
the on-line document \cite{cristia2021log}). This means that \setlog can serve
as a \emph{programming language} in which \emph{prototypes} of those
specifications can be immediately implemented. Then, \setlog itself can be used
to automatically prove that the specifications preserve some state invariants.

%%%%%%%%%%%%%%%%%%%%%%%%%%%%%%%%%%%%%%%%%%%%%%%%%%%%%%%%%%%%%%%%%%%%%
\section{Dealing with Binary Relations and Partial Functions}
%%%%%%%%%%%%%%%%%%%%%%%%%%%%%%%%%%%%%%%%%%%%%%%%%%%%%%%%%%%%%%%%%%%%%

The relational fragment of \setlog is at least as expressive as the class of
full set relation algebras on finite sets
\cite{DBLP:journals/jar/CristiaR20,DBLP:conf/RelMiCS/CristiaR18}. In spite of
the inherent undecidability of this class of relation algebras, \setlog is able
to automatically reason about practical problems expressed in relational terms.
%As such it can express a number of set and relational properties.

\begin{example}
The overriding operator present in the Z formal notation is defined as follows:
\[
R \oplus S = ((\dom S) \ndres R) \cup S
\]
where $R$ and $S$ are binary relations and $\ndres$ is domain
anti-restriction. Given that the operation that updates a table can be modeled
as an overriding operation, $\oplus$ is frequently used in Z specifications.

Overriding is available in \setlog in the form of the (derived) constraint
$\Oplus$:
\[
\Oplus(R,S,T) \iff T = R \oplus S
\]
Hence, we can specify in \setlog the $\mathsf{update}$ operation of Example
\ref{ex:update} as follows:
\begin{gather*}
\mathsf{updateOplus}(F,X,Y,F\_,Error) \text{ :-} \\
\t1 \Oplus(F,\{[X,Y]\},F\_) \And Error = ok \\
\t1 \Or \\
\t1 \Dom(F,D) \And X \Nin D \And Error = err.
\end{gather*}
Then we can use \setlog to prove that $\mathsf{update}$ refines
$\mathsf{updateOplus}$:
\[
\mathsf{update}(F,X,Y,G) \implies \mathsf{updateOplus}(F,X,Y,G)
\]
by proving the negation to be unsatisfiable. In this way we get a more
efficient code given that $\Oplus$ is too powerful when one only wants to
update a single point in the relation. \qed

%here the problem is that we can't automatically compute the negation of
%\textsf{updateOplus} unless we implement the negation algorithm we discussed at
%Parma}
\end{example}

\begin{remark}
Logical negation can be avoided in \setlog as long as we work with primitive
constraints, since for each of them \setlog implements also its negation. On
the other hand, if the formula to be negated is a compound formula (i.e., a
formula formed by conjunction and disjunction of atomic predicates, such as,
for instance, $\mathsf{updateOplus}$ in the above example), then we must
distribute ``by hand'' the negation all the way down to the atoms at which
point we use the negations of the primitive constraints.

Automating the generation of such kind of negated formulas is one of the
improvements that are planned as future work. \qed
\end{remark}

The decidable fragment of the relational fragment of \setlog is still very
expressive. In fact, for a formula to be outside the decision procedure it must
contain an atom such as $\Comp(R,S,\{X/R\})$ or $\Comp(S,R,\{X/R\})$ or a
subformula that in some way hides such atoms, i.e., it must contain a
relational composition where one of the operands shares a variable with the
result of the composition.\footnote{The technical details are more complex but
this is the essence of the problem.} For example, if $\Dom(\{X/R\},A) \And
\Ran(R,A)$ is present in the formula, chances are that it will lay outside the
decision procedure, since $\Dom$ and $\Ran$ constraints are rewritten to
formulas based on $\Comp$. When a formula lays outside the decision procedure
\setlog will enter an infinite loop. This means that \setlog gives correct
answers, but it might not give an answer.

The absence of $\Comp$ constraints of the special form mentioned above is only
a sufficient condition for termination of \setlog. In fact, not all formulas
containing such constraints go into an infinite loop. For example, the formula
$\Comp(\{[X,Z]/R\},\{[Z,Y]/S\},R) \land \Id(A,R)$, where the first and the
third operands share the same variable $R$, terminates returning a finite
number of solutions. Further investigation on the kind of formulas that makes
\setlog to enter an infinite loop is left for future work. For now, we can
observe that these patterns seldom occur in practice. Indeed, an extensive
empirical evaluation of a \setlog shows that the solver is able to
automatically prove hundreds of theorems of set theory and relation algebra on
finite sets, and to automatically find solutions to systems of constraints of
the same theories, as well \cite{DBLP:journals/jar/CristiaR20}.

%%%%%%%%%%%%%%%%%%%%%%%%%%%%%%%%%%%%%%%%%%%%%%%%%%%%%%%%%%%%%%%%%%%%%
\section{Dealing with Set Cardinalities}
%%%%%%%%%%%%%%%%%%%%%%%%%%%%%%%%%%%%%%%%%%%%%%%%%%%%%%%%%%%%%%%%%%%%%

Some times it is necessary to reason about the size of data structures and not
only about their contents. For example, within the algebra of finite sets one
can partition a given set into two disjoint subsets: $C = A \cup B \land A \cap
B = \emptyset$. But there is no way to state that $A$ and $B$ must be of the
same cardinality. In practice, these constraints might appear, for instance,
when part of a given data structure must be put into a cache when its size
reaches certain threshold. Specifically, cardinality constraints appear in the
verification of some distributed algorithms
\cite{DBLP:conf/cav/BerkovitsLLPS19,Alberti2017} and are at the base of the
notions of integer interval, arrays and lists.

\setlog implements a decision procedure for the algebra of finite sets with
cardinality \cite{DBLP:journals/corr/abs-2102-05422}. In this regard \setlog combines the rewrite
rules of the CLP(SET) scheme with a decision algorithm for formulas including
cardinality developed by C. Zarba \cite{DBLP:conf/frocos/Zarba02}. Zarba proves
that a theory of finite sets equipped with the classic set theoretic operators,
including cardinality, combined with linear integer constraints is decidable.
The \setlog decision procedure first uses all the power of \setlog to produce a
simplified, equivalent formula that can be passed to Zarba's algorithm which
makes a final judgment about its satisfiability, in case it contains
cardinality constraints. At implementation level Zarba's algorithm is
implemented by integrating the Prolog Boolean SAT solver developed by Howe and
King \cite{DBLP:journals/tcs/HoweK12} with SWI-Prolog's implementation of the
CLP(Q) system \cite{holzbaur1995ofai}. As a result the implementation
integrates three Prolog-based systems: Howe and King's SAT solver, CLP(Q) and
\setlog.

Hence, \setlog can be used to automatically prove verification conditions based
on the cardinality operator.

\begin{example}
\setlog has been tested against +250 verification conditions arising during the
analysis of distributed algorithms \cite{Piskac2020}. For instance, it can
automatically discharge the following proof obligation.
\begin{gather*}
 \Size(U,N) \And N > 0 \And N > 3*T \And \\
 \Subseteq(F,U) \And \Size(F,M) \And M \leq T \And \\
 \Subseteq(Cgs,U) \And \Size(Cgs,K) \And 2*K \geq N - T + 1 \And \\
 \Subseteq(Bgr,U) \And \Size(Bgr,J) \And 2*J \geq N + 3*T + 1 \And \\
 \Cap(Cgs,Bgr,L) \And \Size(L,0) \tag*{\qed}
\end{gather*}
\end{example}

As a consequence of the fact that the new decision procedure is still based on
set unification, it can deal with set of sets nested at any level. For example,
the decision procedure is able to give all the possible solutions for a goal
such as $\Size(\{\{X\},\{Y,Z\}\},N)$, where $X$, $Y$, $Z$ and $N$ are
variables.

The formulas returned by \setlog represent all the concrete (or ground)
\emph{solutions} of the input formula. If these formulas do not contain any
$\Size$ or integer constraints, then a concrete solution for such formulas is
obtained using the empty set for all set variables occurring in them (with the
exception of the variables $X$ in atoms of the form $X = t$). Unfortunately,
this is no longer true when considering also the $\Size$ and integer
constraints. For example the answer to the following formula:
\begin{gather*}
\Size(A,M) \And 1 \leq M \And M \leq 2 \And \Size(B,N) \And 5 \leq N \And \\
\Subseteq(C,B) \And \Size(C,K) \And 7 \leq K.
% {log}=> size(A,M) & 1 =< M & M =< 2 & size(B,N) & 5 =< N &
%         subset(C,B) & size(C,K) & 7 =< K.
\end{gather*}
is
\begin{gather*}
 \mathsf{true} \\
\begin{split}
 \text{$\mathsf{Constraint:\,}$} & \Size(A,M), M \ge 0, 1 \le M, M \le 2, \Size(B,N), N
 \ge0, \\
 & 5 \le N, \Subseteq(C,B), \Size(C,K), K \ge 0, 7 \le K
% true
% Constraint: size(A,M), M>=0, 1=<M, M=<2, size(B,N), N>=0,
%            5=<N, subset(C,B), size(C,K), K>=0, 7=<K
\end{split}
\end{gather*}
That is, \setlog returns the formula itself. This means the formula is
satisfiable and that all the possible solutions can be obtained by fixing
values for the variables as long as all the constraints are met. However, this
answer does not point out an evident concrete solution for the formula.

For some applications such as model-based testing \cite{CristiaRossiSEFM13}
determining the satisfiability of a formula is not enough. A more or less
concrete solution is needed. For this reason \setlog provides a way in which
the solver returns formulas for which is always easy to find a solution. We
call such a solution a \emph{minimal solution} because the cardinalities of all
the set variables in $\Size$ constraints are the smallest as to satisfy the
formula. When \setlog is executed in the minimal solution mode, the answer to
the above goal is a more concrete solution:
\begin{gather*}
 A = \{\_N8\}, \\
 M = 1, \\
 B = \{\_N7,\_N6,\_N5,\_N4,\_N3,\_N2,\_N1\}, \\
 N = 7, \\
 C = \{\_N7,\_N6,\_N5,\_N4,\_N3,\_N2,\_N1\}, \\
 K = 7 \\
 \text{$\mathsf{Constraint:\,}$} \_N7 \Neq \_N6, \_N7 \Neq \_N5, ..., \_N3 \Neq \_N1, \_N2 \Neq \_N1
% A = {_N8},
% M = 1,
% B = {_N7,_N6,_N5,_N4,_N3,_N2,_N1},
% N = 7,
% C = {_N7,_N6,_N5,_N4,_N3,_N2,_N1},
% K = 7
% Constraint: _N7 neq _N6, _N7 neq _N5, ..., _N3 neq _N1, _N2 neq _N1
\end{gather*}
This formula is a finite representation of a subset of the possible solutions
for the input formula from which it is trivial to get concrete solutions.

%%%%%%%%%%%%%%%%%%%%%%%%%%%%%%%%%%%%%%%%%%%%%%%%%%%%%%%%%%%%%%%%%%%%%
\section{Restricted Intensional Sets}
%%%%%%%%%%%%%%%%%%%%%%%%%%%%%%%%%%%%%%%%%%%%%%%%%%%%%%%%%%%%%%%%%%%%%

%RIS can be used in many ways as a powerful programming and reasoning concept.
Intensional sets are widely recognized as a key feature to describe complex
problems, possibly leading to more readable and compact programs than those
based on conventional data abstractions. As a matter of fact, various
specification or modeling languages provide intensional sets as first-class
entities.

\setlog provides a narrower form of intensional sets, called Restricted
Intensional Sets (RIS), that are similar to the set comprehensions available in
the formal specification language Z. The basic form of a RIS term is:
\[
\Ris(X \In A, \flt, \ptt)
\]
where $D$ is a set, $\flt$ is a \setlog formula, and $\ptt$ is a term
containing $X$. The intuitive semantics of this RIS is ``the set of instances
of the term $\ptt(X)$ such that $X$ belongs to $D$ and $\flt$ holds for $X$'',
i.e., $\{y : \exists x(x \in A \land \flt \land y = \ptt(x)\}$.

RIS have the restriction that $D$ must be a \emph{finite set}. This fact, along
with a few restrictions on variables occurring in $\flt$ and $\ptt$, guarantees
that the RIS is a finite set, given that it is at most as large as $D$. It is
important to note that, although RIS are guaranteed to denote finite sets,
nonetheless, RIS may be not completely specified. In particular, as the domain
can be a variable or a partially specified set, RIS are finite but
\emph{unbounded}.

\setlog formulas containing RIS remain decidable if the formulas inside them
are decidable\footnote{Among others, more technical, restrictions
\cite{DBLP:conf/cade/CristiaR17,Cristia2021}.}.

The next example shows the classes of problems RIS are meant to solve.

\begin{example}\label{ex:first}
First, we can use \setlog with RIS as a \emph{programming language}. We can
think in a program outputting the even numbers ($E$) present in a set of
numbers ($S$) that is the input to the program:
\begin{equation}\label{eq:ex1prog}
%E = \{x:S | x \mod 2 = 0 @ x\}
E = \Ris(X \In S, x \mod 2 = 0, X)
\end{equation}
The RIS term can be written more compactly as $\Ris(X \In S, x \mod 2 = 0)$,
since in this case its third argument coincides with its control variable
(i.e., the first argument). Then if $S$ is bound to $[-2,2]$, \setlog will
answer $E = \{-2,0,2\}$.

Second, we can use \setlog with RIS as a \emph{solver} for set formulas. For
instance, we want \setlog to find the most general solution for the following
formula:
\[
\Ris(X \In A, 0 \:\mathsf{is}\: X \mod 2) = \{-2,0,2\}
%\{-2,0,2\} = \{x:S | x \mod 2 = 0\}
\]
Note that, in a sense, we are asking \setlog to find the input values that make
program \eqref{eq:ex1prog} to return a given output. In this case the answer
will be:
\[
%S = \{-2,0,2 \plus N\} \land \{x : N | x \mod 2 = 0\} = \es
S = \{-2,0,2 / N\}, \Ris(X \In N, X \mod 2 = 0\} = \es
\]
where $N$ is a new variable (implicitly existentially quantified). Substituting
$N$ by $\es$ yields a ground solution (i.e., $S = \{-2,0,2\}$).

Third, we can use \setlog with RIS to \emph{prove} properties of \setlog
formulas.
%In this case we use $\setlog$ as a satisfiability solver and theorem prover:
%
%Then, if we want to prove a property we call \setlog on the negation of the
%formula expecting that it answers $\false$.
For instance, to prove that $\Cup(\{x:S | x < m\},\{x:S | x > m\},B) \implies m
\notin B$, we can prove the following \setlog formula:
\[
%\Cup(\{x:S | x < m\},\{x:S | x > m\},B) \land m \in B
\Cup(\Ris(X \In S, X < M),\Ris(X \In S, X > M),B) \land M \In B
\]
which is found to be unsatisfiable. \qed
\end{example}

%%%%%%%%%%%%%%%%%%%%%%%%%%%%%%%%%%%%%%%%%%%%%%%%%%%%%%%%%%%%%%%%%%%%%%%%%%%%%%
\begin{comment}
Another important use of RIS is to define \emph{(partial) functions} by giving
their domains and the expressions that define them. Given that RIS are sets,
then functions are sets of ordered pairs as in Z and B. Therefore, through
standard set operators, functions can be evaluated, compared and point-wise
composed; and by means of constraint solving, the inverse of a function can
also be computed.

\begin{example}
A function that maps integers belonging to a set $D$ to their squares. Using an
abstract notation for RIS):
\[
\{x:D | \true @ (x,x*x)\}
\]
The square of 5 can be calculated by: $(5,y) \in \{x:D @ (x,x*x)\}$, yielding
$y = 25$. The same RIS calculates the square root of a given number: $(x,36)
\in \{x:D @ (x,x*x)\}$, returning $x = 6$ and $x = -6$. Set membership can also
be used for the point-wise composition of functions. The function $f(x) = x^2 +
8$ can be evaluated on 5 as follows: $(5,y) \in \{x:D @ (x,x*x)\} \land (y,z)
\in \{e:E @ (e,e + 8)\}$ returning $y = 25$ and $z = 33$. \qed
\end{example}
\end{comment}
%%%%%%%%%%%%%%%%%%%%%%%%%%%%%%%%%%%%%%%%%%%%%%%%%%%%%%%%%%%%%%%%%%%%%%%%%%%%%%

As \textsf{LIA} is decidable, RIS are a convenient mechanism to model and
reason about programs dealing with integers.

\begin{example}\label{ex:RIS-ILA}
RIS can be used to get the subset of a set verifying some \textsf{LIA} formula,
which cannot be done rather efficiently in a pure algebraic fragment of set
theory.
\begin{gather*}
\Ris(X \In S, \Integer(X)) \ie{$S \cap \num$} \\
\Ris(X \In S, M \leq X \And X \leq N) \ie{$S \cap [M,N]$} \\
\Ris(X \In S, 0 \leq X) \ie{$S \cap \nat$} \\
\Ris(X \In S, 0 \leq 3*X + 2*Y) \tag*{\qed}
\end{gather*}
\end{example}

The same can be done with binary relations. As an example, $\Ris([X,Y] \In R, X
\leq Y)$ represents the binary relation $R \cap (\_\leq\_)$.

%\begin{gather*}
%\Ris([X,Y] \In R, X \leq Y) \ie{$R \cap (\_\leq\_)$} \\
%\Ris([X,Y] \In R, Y \Is 2*X + 3) \ie{$R \cap \text{a rect line}$}\\
%\Ris([X,Y] \In R, Y \Is -X) \ie{$R \cap abs$} \\
%\Ris([[V,W],[X,Y]] \In R, V \leq X \implies W \leq Y) \tag*{\qed}
%\end{gather*}

\begin{remark}

The language of RIS, called $\LRIS$, is parametric with respect to any
first-order theory $\TX$ providing at least equality and a decision procedure
for $\TX$-formulas.
%all of the theoretical results for the RIS theory apply provided RIS' filters
%belong to a decidable \emph{fragment} of the considered parameter theory, and a
%solver for this fragment is called from $\SATRIS$.
In practice, however, many interesting theories are undecidable and only
semi-decision procedures exist for them. This is the case, for instance, for
the theory \textsf{RA} of sets and binary relations implemented by \setlog.
Hence, the condition on the availability of a decision procedure for \emph{all}
$\TX$-formulas can be often relaxed. Instead, the existence of some
algorithm capable of deciding the satisfiability of a significant fragment of
$\TX$-formulas can be assumed. If such an algorithm exists and the user
writes formulas inside the corresponding fragment, all the theoretical results
for RIS still apply. \qed
%For these reasons, in coming sections, we sometimes give examples where $\Ur$
%is not necessarily a decidable theory.
\end{remark}

%%%%%%%%%%%%%%%%%%%%%%%%%%%%%%%%%%%%%%%%%%%%%%%%%%%%%%%%%%%%%%%%%%%%%
\section{Universal Quantification in \setlog}
%%%%%%%%%%%%%%%%%%%%%%%%%%%%%%%%%%%%%%%%%%%%%%%%%%%%%%%%%%%%%%%%%%%%%

Formulas that \setlog can deal with are quantifier-free first-order formulas
over finite sets and integer linear arithmetic.
%over a significant fragment of classical set-theory, extended with binary
%relations and integer linear arithmetic
%\setlog provides an effective semi-procedure \cite[Def. 1.5.13]{hinman2018fundamentals}
%for dealing with such formulas.

However, \setlog provides also some form of universal quantification by means of
RIS. In effect, the introduction of RIS in \setlog allows for the definition of
\emph{restricted universal quantifiers} (RUQ). In general, if $A$ is a set,
then a RUQ is a formula of the following form:
\[
\forall x \in A: \phi
\]
It is easy to prove the following:
\begin{equation}\label{e:prop1}
(\forall x \in A: \phi) \iff A \subseteq \{x : x \in A \land \phi\}
\end{equation}
Given that $\{x : x \in A \land \phi\}$ is the interpretation  of $\Ris(X \In
A,\phi)$, the r.h.s. of \eqref{e:prop1} can be expressed as the \setlog
formula:
\[
\Subseteq(A,\Ris(X \In A,\phi))
\]
for which \setlog provides the derived constraint $\Foreach$ thus making RUQ
easier to write:
\begin{equation}\label{eq:foreach1}
\Foreach(X \In A,\phi) \defs \Subseteq(A,\Ris(X \In A,\phi)).
\end{equation}

There is also a more powerful form of $\Foreach$:
\begin{equation}\label{eq:foreach2}
\Foreach(X \In A,\mathsf{V},\phi,\psi) \defs \Subseteq(A,\Ris(X \In
A,\mathsf{V},\phi,\psi)).
\end{equation}
where $\mathsf{V}$ is a vector of existentially quantified variables inside the
$\Foreach$ and $\psi$ is a so-called \emph{functional predicate}. A predicate
$\psi(x_1,\dots,x_{n+1})$ is a functional predicate iff for any given
$x_1,\dots,x_n$ there exists at most one $x_{n+1}$ making $\psi$ true.
Functional predicates enjoy a nice property concerning their negation
\cite{jar-ris}, which considerably extends the class of decidable formulas
including $\Foreach$ constraints.

\begin{example}\label{ex:seccond}
We use $\Foreach$ to encode and automatically reason about important security
properties \cite{DBLP:journals/jar/CristiaR21}. The Bell-LaPadula (BLP) security model proposes
two security properties for secure operating systems. The simplest is called
security condition of which we show a simplified version\footnote{Just look at
the complexity of the formula, not its interpretation in terms of computer
security.}:
\newcommand{\re}{\mathsf{r}}
\newcommand{\w}{\mathsf{w}}
\newcommand{\ap}{\mathsf{a}}
\newcommand{\co}{\mathsf{c}}
\begin{equation}\label{eq:seccond}
\forall (s,o,x) \in b:
x = \re \implies
       f_2(o) \leq f_1(s) \land f_4(o) \subseteq f_3(s)
\end{equation}
where $b$ is a ternary relation and $f_i$ are functions.  In \setlog we can
encode a ternary relation with a binary relation where the second components
are ordered pairs. Then $(s,o,x) \in b$ becomes $[S,[O,X]] \In B$. Therefore,
\eqref{eq:seccond} is encoded as follows:
\begin{gather*}
\mathsf{seccond}(F1,F2,F3,F4,B) \text{ :- } \\
\t1 \Foreach([S,[O,X]] \In B,[No,Co,Ns,Cs], \\
\t2 No \leq Ns \And \Subseteq(Co,Cs), \\
\t2 \Apply(F1,S,Ns) \And \Apply(F2,O,No) \And \\
\t2 \Apply(F3,S,Cs) \And \Apply(F4,O,Co))
\end{gather*}
where $\Apply$ is a derived constraint which is defined as $\Apply(F,X,Y) \defs
[X,Y] \In F \And \Pfun(F)$. Note that all the $\Apply$ constraints are placed
in the last argument given that they are functional predicates. \qed
\end{example}

\begin{remark}
Example \ref{ex:seccond} brings in the point of when a binary relation can be
\emph{applied} to an element of its domain. Usually the condition for function
application is, precisely, for the binary relation to be a function. However,
there is a weaker condition in which the binary relation is \emph{locally}
functional;  that is, the binary relation is a function in (at least) one
point. Therefore, in \setlog the user can work also with the following derived
constraint:
\[
\ApplyTo(F,X,Y) \defs \\
    \t1 F = \{[X,Y] / G\} \And [X,Y] \Nin G \And \Comp(\{[X,X]\},G,\{\}).
\]
Using $\Apply$ as in Example \ref{ex:seccond} implies that $\mathsf{seccond}$
checks that $F1$-$F4$ are functions every time it is called. In a real
implementation this is not the case because the fact that $F1$-$F4$ are
functions would be a pre-condition.  Hence, a more realistic specification
would use $\ApplyTo$ instead of $\Apply$. We can use \setlog to automatically
prove that if $F$ is a function then $\ApplyTo$ refines $\Apply$
\[
\Pfun(F) \And \ApplyTo(F,X,Y) \implies \Apply(F,X,Y)
\]
This allows to formally replace $\Apply$ by $\ApplyTo$ in $\mathsf{seccond}$.
Moreover, this results in a more efficient implementation as $\ApplyTo$ is
linear in the size of $F$ while $\Apply$ is quadratic. As a matter of fact,
discharging all the verification conditions of the BLP model using $\ApplyTo$
is almost 10 times faster than when using $\Apply$. \qed
\end{remark}

The $\Foreach$ constraint can be used to model and reason about order.  In fact
if $F$ is a function with domain and range in $\num$, then we can use \setlog
to define a predicate stating whether or not $F$ is a strictly increasing
injective function ($\mathsf{siif}$).

\begin{example}
The following formula captures the notion of
strictly increasing function $F$:
\begin{gather*}
  \forall(a,c) \in F,\forall(x,y) \in F : a < x \implies c < y
\end{gather*}
which is immediately rendered in \setlog by the following predicate:
\begin{gather*}
\mathsf{siif}(F) \text{ :- } \\
  \t1 \Pfun(F) \\
  \t1 \land \Foreach([A,C] \In F, \\
  \t1 \t1 \Foreach([X,Y] \In F, A \ge X \Or C < Y)))  \tag*{\qed}
\end{gather*}
\end{example}

%\begin{remark}
Although $\Foreach$ can be defined as a derived constraint based on the
constraint $\Cup$, in \setlog we introduce a set of specialized rewrite rules
to process this specific kind of predicates more efficiently \cite{jar-ris}.
As a matter of fact, the formula $\Foreach(x \in \set{t}{A},\flt(x))$ can be
seen as an iterative program whose iteration variable is $x$, the range of
iteration is $\set{t}{A}$, and the body is $\flt$. In fact, the rewrite rule
for this formula basically iterates over $\set{t}{A}$ and evaluates $\flt$ for
each element in that set. If one of these elements does not satisfy $\flt$ then
the loop terminates immediately, otherwise it continues until the empty set is
found or a variable is found.
%\end{remark}

%%%%%%%%%%%%%%%%%%%%%%%%%%%%%%%%%%%%%%%%%%%%%%%%%%%%%%%%%%%%%%%%%%%%%
\section{Finite Integer Intervals}
%%%%%%%%%%%%%%%%%%%%%%%%%%%%%%%%%%%%%%%%%%%%%%%%%%%%%%%%%%%%%%%%%%%%%

The theory \INT (cf. Figures \ref{fig:stack} and \ref{fig:theories})  deals
with hereditarily finite hybrid untyped extensional sets and integer intervals
with cardinality. The bounds of integer intervals can be either integer
constants or variables ranging over integer numbers, and as such they can be
constrained through integer linear arithmetic constraints.

\setlog provides a decision procedure of the theory \INT. Integer intervals in
\setlog are represented by terms of the form $\Int(A,B)$, where $A$ and $B$ are
integer constants or variables, which is interpreted as the close interval
$[A,B]$. Interval terms can be manipulated as sets through set constraints
(e.g., $\Int(0,B) = \{1,2/S\}$ or $\Cap(\Int(3,N),\Int(10,20),A) \And A \Neq
\es$). Moreover, interval bounds can be manipulated as integers through integer
constraints (e.g., $A = \Int(M,N) \And N >= M + 3$).

The decision procedure for the theory \INT allows \setlog to be used to program
and automatically reason about problems such as the following.
\begin{example}
Consider two workers who are assigned two disjoint sets of tasks from  a set of
$N$ tasks. If $A$ and $B$ are the sets of tasks already performed by each
worker, then we can model the problem as follows.
\begin{gather*}
\mathsf{init}(A,B) \text{ :- } \\
\t1  A = \{\} \And B = \{\}. \\
\mathsf{addToA}(N,A,B,J,A\_,B) \text{ :- } \\
\t1 (J \Nin \Int(1,N) \Or J \In A \Or J \In B) \And
    A\_ = A \\
\t1   \Or \\
\t1   J \In \Int(1,N) \And J \Nin A \And J \Nin B \And
      A\_ = \{J / A\}. \\
\mathsf{finish}(N,A,B,\{\},\{\}) \text{ :- } \\
\t1  \Cup(A,B,\Int(1,N)) \And \mathsf{write('Job done')}. \\
\mathsf{\%\ Invariant 1: } \Cup(A,B,C) \And \Subseteq(C,\Int(1,N)) \\
\mathsf{\%\ Invariant 2: } \Disj(A,B)
\end{gather*}
For brevity we do not show $\mathsf{addToB}$
which would symmetric to $\mathsf{addToA}$. Then, we can use \setlog to run
some simulations:
\begin{gather*}
\mathsf{init}(A,B) \And \mathsf{addToA}(10,A,B,3,A1,B1) \And
\mathsf{addToB}(10,A1,B1,7,A2,B2).
\end{gather*}
Finally we can use \setlog to automatically prove the indicated invariants.

If $A$ and $B$ must done $N/2$ tasks each, then we can add a $\Size$-based
pre-condition to $\mathsf{addToA}$  and $\mathsf{addToB}$, which would be still
inside the decision procedures implemented by \setlog. \qed
\end{example}

\begin{example}
Assume some objects are identified with numbers from 1 up.  We want to write a
condition stating that a certain set of these objects, $A$, contains objects
with consecutive numbers. It can be written with an \INT-based formula:
\[
A = \Int(M,N) \And 1 \leq M
\]
Moreover, if $A$ does not verify that condition we would like to  compute the
missing objects from it:
\[
A = \Int(M,N) \And 1 \leq M \\
\Or \\
\Nint(A) \And \mathsf{min}(A,M) \And \mathsf{max}(A,N) \And
\Cup(A,Miss,\Int(M,N))
\]
where $\Nint$ is a predicate stating that $A$ is not an integer interval,  and
$\mathsf{min}$ and $\mathsf{max}$ compute the minimum and maximum of $A$; all
of which can be stated as \INT formulas. \qed
\end{example}

The key idea for obtaining a decision procedure for the theory \INT is
extending the set unification algorithm of \CLPSET \cite{Dovier00} with the
following identity:
\begin{equation*}
A = [m,n] \iff A \subseteq [m,n] \land \card{A} = n-m+1
\end{equation*}
In fact, it suffices to be able to deal with constraints of the form $A
\subseteq [m,n]$ in a decidable manner to have a decision procedure for integer
intervals.

Exploiting extended set unification with intervals \setlog allows, for
instance, to reconstruct integer intervals even out of underspecified sets:
\[
\Cup(\{X,Y,Z\},\{1,4\},\Int(M,N)) \\
% un({X,Y,Z},{1,4},int(M,N))
\]
of which some solutions are:
\[
X = 0,
Y = 2,
Z = 3,
M = 0,
N = 4 \\
X = 2,
Y = 3,
Z = 5,
M = 1,
N = 5 \\
X = 2,
Y = 3,
Z = 1,
M = 1,
N = 4
\]

\begin{remark}
In some cases there are a few different ways of writing the same term or
formula. For instance, the RIS term $\Ris(X \In S, M \leq X \And X \leq N)$ of
Example \ref{ex:RIS-ILA} can be written as the \INT formula
$\Cap(S,\Int(M,N),A)$, in which case we have $A = \Ris(X \In S, M \leq X \And X
\leq N)$. Which is the best language construct to express programs and
properties depends on, some times, contradictory concepts such as efficiency
and readability. At least \setlog provides a way to go from one construct to
another. That is, if a user writes a formula containing the RIS in question,
(s)he can substitute the RIS by $A$ if $\Cap(S,\Int(M,N),A)$ is conjoined to
the formula, after using \setlog to prove the substitution is correct. \qed
% ninters(S,int(M,N),ris(X in S, M =< X & X =< N)).
\end{remark}

%\subsection{Intervals vs. quantifiers}
It is worth noting that a combination between the subset relation and integer
intervals is the key to encode forms of universal quantification in \setlog by
means of a quantifier-free formula, allowing us to preserve decidability (and
thus full automation in proofs).
%As the above examples show, a key aspect to preserve decidability (and thus
%full automation in proofs) is not to use logical negation but the negated
%constraints provided by \setlog.  In turn, this is possible because some
%constraints implement different forms of universal quantification.
The following example illustrates this idea.

\begin{example}
If $X, Y \in D$, $Y$ is the successor of $X$ (in $D$) if the following holds:
\[
\lnot \exists Z \in D: X < Z \land Z < Y
\]
which is equivalent to:
\begin{equation}\label{eq:suc}
\forall Z \in D: Z \leq X \lor Y \leq Z
\end{equation}
In this case we need to quantify over integer numbers. A way to get rid of this
universal quantifier (hence, obtaining a quantifier-free \setlog formula) is to
use a combination between the subset relation and integer intervals as follows:
\begin{gather*}
\mathsf{succ}(D,X,Y)  \text{ :-} \\
\t1 D = \{X,Y / E\} \And \Cup(Inf,Sup,E) \And \Disj(Inf,Sup) \And\\
\t1 M \Is X-1 \And \Subseteq(Inf,\Int(\_,M)) \And \\
\t1 N \Is Y+1 \And \Subseteq(Sup,\Int(N,\_)).
\end{gather*}
To confirm that $\mathsf{succ}$ is indeed an encoding of \eqref{eq:suc} we can
execute some tests:
\begin{gather*}
\mathsf{succ}(\{4,7,1,8,-3\},1,Y) \rightarrow Y = 4 \\
\mathsf{succ}(\{4,7,1,8,-3\},2,Y) \rightarrow \mathsf{no}
   \why{$2 \notin D$} \\
\mathsf{succ}(\{4,7,1,8,-3\},8,Y) \rightarrow \mathsf{no}
   \why{$max(D) = 8$}\\
\mathsf{succ}(\{4,7,1,8,-3\},4,Y) \rightarrow Y = 7 \\
\mathsf{succ}(\{4,7,1,8,-3\},X,7) \rightarrow X = 4
   \tag{$\dagger$} \label{succ:test1}\\
\mathsf{succ}(\{4,7,1,8,-3\},X,Y) \rightarrow
   X = 4, Y = 7; X = 7, Y = 8; \dots
   \tag{$\ddagger$} \label{succ:test2}
\end{gather*}
Note that \eqref{succ:test1} shows that \setlog  does not really distinguish
between inputs and outputs; and \eqref{succ:test2} shows that \setlog is able
to return all solutions one after the other. Furthermore, to collect stronger
evidences that $\mathsf{succ}$ is correct we can use \setlog to automatically
prove some properties true of it:
\[
\mathsf{succ}(D,X,Y) \implies (\forall Z \in D: Z \leq X \lor Y \leq Z)
\]
whose negation is:
\[
\mathsf{succ}(D,X,Y) \And Z \In D \And X < Z \And Z < Y
\]
to which \setlog answers $\mathsf{no}$. And further we can prove:
\[
\mathsf{succ}(D,X,Y) \And \mathsf{succ}(D,Y,Z) \implies X < Z
\]
whose negation is:
\begin{equation*}
\mathsf{succ}(D,X,Y) \And \mathsf{succ}(D,Y,Z) \And Z \leq X
   \tag*{\qed}
\end{equation*}
\end{example}

Finally, integer intervals are a key component in the definition of arrays and
list as sets; hence, to implement the theories \textsf{ARRAY} and \textsf{LIST}
in \setlog (cf. Figure \ref{fig:stack}). In particular, if $\mathsf{array}(A,n)$ is a predicate stating that
$A$ is an array of length $n$ whose components take values on some universe
$\Ur$, then it can be defined as follows:
\begin{equation*}
array(A,n) \iff A: [1,n] \fun \Ur
\end{equation*}
i.e., as a partial function between the integer interval $[1,n]$ and $\Ur$.
Since \setlog supports a broad class of set relation algebras, including
partial functions and the domain operator, then it would be possible to use
\setlog to automatically reason about broad classes of programs with arrays.
Lists could be introduced in a similar way.

However, supporting arrays and lists in \setlog is a line of future research.
%is left for future work

%%%%%%%%%%%%%%%%%%%%%%%%%%%%%%%%%%%%%%%%%%%%%%%%%%%%%%%%%%%%%%%%%%%%%
\section{Concluding Remarks}\label{concl}
%%%%%%%%%%%%%%%%%%%%%%%%%%%%%%%%%%%%%%%%%%%%%%%%%%%%%%%%%%%%%%%%%%%%%

The CLP language \setlog provides decision procedures for expressive classes of
extensional and intensional hereditarily finite hybrid sets, including binary
relations, integer intervals and Cartesian products, extended with cardinality
constraints and integer constraints for integer linear arithmetic.

In this paper we have shown how \setlog, with its decision procedures, can be
exploited: $(i)$ as a \emph{programming language}, in which a \emph{prototype}
of a set-based specification can be immediately implemented; $(ii$) as a
\emph{satisfiability solver} for formulas of the different theories, in
particular for formulas representing the implementation of a set-based
specification for which \setlog can be used to \emph{prove} that certain
properties are true of the specification.
%We have shown this ``double'' use of \setlog through
In this paper we have provided evidence for this claim by showing a number of
simple working examples written in \setlog.

Besides the possible future work pointed out throughout the paper, we are
investigating the possibility to add interactive theorem proving capabilities
to \setlog \cite{setlog-itp-cj} in order to make it capable of proving
properties outside of the implemented decision procedures.

%\nocite{*}
\bibliographystyle{alpha}
\bibliography{/home/mcristia/escritos/biblio.bib}
%%%%%%%%%%%%%%%%%%%%%%%%%%%%%%%%%%%%%%%%%%%%%%

\end{document}